# State/observable interactions using basic geometric algebra solutions of the Maxwell equation

Alexander SOIGUINE[1]

[1] SOiGUINE Quantum Computing, Aliso Viejo, CA 92656, USA

http://soiguine.com

Email address: alex@soiguine.com

**Abstract:** Maxwell equation in geometric algebra formalism with equally weighted basic solutions is subjected to continuously acting Clifford translation. The received states, operators acting on observables, are analyzed with different values of the Clifford translation time factor and through the observable measurement results.

## 1. Introduction

Let's consider special case of g-qubits [1] [2]:

$$\alpha + \beta B_1 + \alpha B_2 + \beta B_3 = \alpha + \beta B_3 + (\alpha + \beta B_3)B_2 \qquad (1.1)$$

with basic bivectors satisfying usual anticommutative multiplication rules: $B_1 B_2 = -B_3$, $B_1 B_3 = B_2$, $B_2 B_3 = -B_1$. The normalization in this case is:

$$\alpha^2 + \beta^2 + \alpha^2 + \beta^2 = 1 \Longrightarrow \alpha^2 + \beta^2 = \tfrac{1}{2} \qquad \text{or} \qquad (\sqrt{2}\alpha)^2 + (\sqrt{2}\beta)^2 = 1$$

With this normalization $\sqrt{2}\alpha = \cos\varphi$ for some angle $\varphi$. Then $\beta = \tfrac{1}{\sqrt{2}}\sin\varphi$, $\varphi = \cos^{-1}(\sqrt{2}\alpha)$. The state (1.1) is one possible lift of conventional quantum mechanical qubit $\begin{pmatrix} \alpha+i\beta \\ \alpha+i\beta \end{pmatrix}$ with $\alpha^2 + \beta^2 = \tfrac{1}{2}$, when complex plane is specified as the plane of $B_3$.

Write state (1.1) in exponential form:

$$\alpha + \beta B_1 + \alpha B_2 + \beta B_3 = e^{I_B \varphi}, \qquad (1.2)$$

$$\varphi = \cos^{-1}(\sqrt{2}\alpha), \ I_B = \tfrac{\beta}{\sqrt{1-2\alpha^2}} B_1 + \tfrac{\alpha}{\sqrt{1-2\alpha^2}} B_2 + \tfrac{\beta}{\sqrt{1-2\alpha^2}} B_3$$

Apply Clifford translation in the plane of $B_3$ to (1.2) (see, for example [3]):

$$e^{I_{B_3}\gamma} e^{I_B \varphi} = \cos\gamma \cos\varphi + \sin\gamma \sin\varphi \, (I_{B_3} \cdot I_B)$$

$$+ \sin\gamma \cos\varphi \, I_{B_3} + \cos\gamma \sin\varphi \, I_B + \sin\gamma \sin\varphi \, I_{B_3} \wedge I_B =$$

$$\cos \gamma \cos \varphi - \frac{\beta}{\sqrt{1-2\alpha^2}}\sin \gamma \sin \varphi + \sin \gamma \cos \varphi \, B_3 + \frac{\beta}{\sqrt{1-2\alpha^2}}\cos \gamma \sin \varphi \, B_1$$
$$+ \frac{\alpha}{\sqrt{1-2\alpha^2}}\cos \gamma \sin \varphi \, B_2 + \frac{\beta}{\sqrt{1-2\alpha^2}}\cos \gamma \sin \varphi \, B_3 + \frac{\beta}{\sqrt{1-2\alpha^2}}\sin \gamma \sin \varphi \, B_2$$
$$+ \frac{\alpha}{\sqrt{1-2\alpha^2}}\sin \gamma \sin \varphi \, B_1 =$$

$$\cos \gamma \cos \varphi - \frac{\beta}{\sqrt{1-2\alpha^2}}\sin \gamma \sin \varphi + \left(\frac{\beta}{\sqrt{1-2\alpha^2}}\cos \gamma \sin \varphi + \frac{\alpha}{\sqrt{1-2\alpha^2}}\sin \gamma \sin \varphi\right) B_1$$
$$+ \left(\frac{\alpha}{\sqrt{1-2\alpha^2}}\cos \gamma \sin \varphi + \frac{\beta}{\sqrt{1-2\alpha^2}}\sin \gamma \sin \varphi\right) B_2$$
$$+ \left(\sin \gamma \cos \varphi + \frac{\beta}{\sqrt{1-2\alpha^2}}\cos \gamma \sin \varphi\right) B_3$$

This will be used later for the g-qubits generated by Maxwell equation.

To make the following text more comprehensive let's briefly recall how the system of the electromagnetic Maxwell equations is formulated as one equation in geometric algebra terms [4].

Take geometric algebra element of the form: $F = \vec{E} + I_3\vec{H}$, where $I_3$ is righthand screw unit volume in the three dimensions. The electromagnetic field $F$ is created by some given distribution of charges and currents, also written as geometric algebra multivector: $J \equiv \rho - \vec{j}$. Apply operator $\partial_t + \nabla$ to $F$. The result will be:

$$(\partial_t + \nabla)F = \underbrace{\nabla \cdot \vec{E}}_{scalar} + \underbrace{\partial_t\vec{E} + I_3(\nabla \wedge \vec{H})}_{vector} + \underbrace{\nabla \wedge \vec{E} + I_3\partial_t\vec{H}}_{bivector} + \underbrace{I_3(\nabla \cdot \vec{H})}_{pseudoscalar}$$

Comparing component wise $(\partial_t + \nabla)F$ and $J$ we get:

$$\begin{cases} \nabla \cdot \vec{E} \equiv div\vec{E} = \rho \\ \partial_t\vec{E} + I_3(\nabla \wedge \vec{H}) \equiv \partial_t\vec{E} - rot\vec{H} = -\vec{j} \\ \nabla \wedge \vec{E} + I_3\partial_t\vec{H} \equiv I_3 rot\vec{E} + I_3\partial_t\vec{H} = 0 \\ I_3(\nabla \cdot \vec{H}) \equiv I_3(div\vec{H}) = 0 \end{cases}$$

Thus, we have usual system of Maxwell equations:

$$\begin{cases} div\vec{E} = \rho \\ \partial_t\vec{E} - rot\vec{H} = -\vec{j} \\ \partial_t\vec{H} + rot\vec{E} = 0 \\ div\vec{H} = 0 \end{cases}$$

equivalent to one equation $(\partial_t + \nabla)F = J$

Without charges and currents the equation becomes $(\partial_t + \nabla)F = 0$

Arbitrary linear combination of the two basic solutions of the above Maxwell equation in geometric algebra terms is [5]:

$$\lambda e^{I_{Plane}^+ \varphi^+} + \mu e^{I_{Plane}^- \varphi^-} \quad (1.3)$$

where

$$\varphi^{\pm} = \cos^{-1}\left(\frac{1}{\sqrt{2}} \cos \omega(t \mp [(I_3 I_S) \cdot \vec{r}])\right),$$

$$I_{Plane}^{\pm} = I_S \frac{\sin \omega(t \mp [(I_3 I_S) \cdot \vec{r}])}{\sqrt{1 + \sin^2 \omega(t \mp [(I_3 I_S) \cdot \vec{r}])}} + I_{B_0} \frac{\cos \omega(t \mp [(I_3 I_S) \cdot \vec{r}])}{\sqrt{1 + \sin^2 \omega(t \mp [(I_3 I_S) \cdot \vec{r}])}}$$
$$+ I_{E_0} \frac{\sin \omega(t \mp [(I_3 I_S) \cdot \vec{r}])}{\sqrt{1 + \sin^2 \omega(t \mp [(I_3 I_S) \cdot \vec{r}])}}$$

The triple of unit value basis orthonormal bivectors $\{I_S, I_{B_0}, I_{E_0}\}$ is comprised of $I_S$ bivector, dual, that's received by applying righthand screw unit volume $I_3$, to the propagation direction vector, $I_{B_0}$ is dual to initial vector of magnetic field, $I_{E_0}$ is dual to initial vector of electric field.

The expression (1.3) is linear combination of two geometric algebra states, g-qubits [1], [2], and can particularly be transformed by a Clifford translation, geometric algebra lift of matrix Hamiltonian action on two-dimensional complex vectors, qubits in terms of conventional quantum mechanics.

Suppose conventional matrix Hamiltonian is $\begin{pmatrix} 0 & 2i\gamma \\ -2i\gamma & 0 \end{pmatrix}$. Its geometric algebra lift makes rotation of the state (1.3) in the $I_{E_0}$ plane by angle $2\gamma$[1] and the lift is $e^{2\gamma I_{E_0}}$.

One interesting example of the current formalism is calculating the Berry potential associated with (1.3) state transformation $e^{2\gamma I_{E_0}}\left(\lambda e^{I_{Plane}^+ \varphi^+} + \mu e^{I_{Plane}^- \varphi^-}\right)$:

$$A(\gamma) = \left(\lambda e^{-I_{Plane}^+ \varphi^+} + \mu e^{-I_{Plane}^- \varphi^-}\right) e^{-2\gamma I_{E_0}} I_{E_0} \frac{\partial}{\partial \gamma} e^{2\gamma I_{E_0}} \left(\lambda e^{I_{Plane}^+ \varphi^+} + \mu e^{I_{Plane}^- \varphi^-}\right)$$

$$= \left(\lambda e^{-I_{Plane}^+ \varphi^+} + \mu e^{-I_{Plane}^- \varphi^-}\right) e^{-2\gamma I_{E_0}} (-2 e^{2\gamma I_{E_0}}) \left(\lambda e^{I_{Plane}^+ \varphi^+} + \mu e^{I_{Plane}^- \varphi^-}\right)$$

$$= (-2)\left(\lambda^2 + \mu^2 + \lambda\mu\left(e^{-I_{Plane}^+ \varphi^+} e^{I_{Plane}^- \varphi^-} + e^{-I_{Plane}^- \varphi^-} e^{I_{Plane}^+ \varphi^+}\right)\right)$$

If the two basic solutions are equally weighted, $\lambda = \mu$, we get a potential, instantly nonlocally spread in three-dimensional space and independent of time, and, up to electric/magnetic field amplitude value, the potential is[2]:

---

[1] The same result takes place if Clifford translation makes rotation in the plane of $I_{B_0}$

[2] Contrary to conventional quantum mechanics formalism where states, particularly qubits, are only defined up to a phase, the geometric algebra with variable complex plane paradigm is much deeper theory, thus the statement that Berry potential is not observable due to gauge redundancy makes here no sense

$$A(\gamma)|_{\lambda=\mu} \sim -\cos^2[(I_3 I_S) \cdot \vec{r}]$$

It is scalar field, thus invariant in any $G_3^+$ measurements [1].

## 2. Clifford translation continuously acting on a state received as the Maxwell equation solution

Let's initially calculate $\lambda e^{I_{Plane}^+ \varphi^+} + \mu e^{I_{Plane}^- \varphi^-}$ assuming again $\lambda = \mu$, say both are equal to 1.

$$\begin{aligned}
& e^{I_{Plane}^+ \varphi^+} + e^{I_{Plane}^- \varphi^-} \\
&= \frac{1}{\sqrt{2}} \cos \omega(t - [(I_3 I_S) \cdot \vec{r}]) + \frac{1}{\sqrt{2}} I_S \sin \omega(t - [(I_3 I_S) \cdot \vec{r}]) \\
&+ \frac{1}{\sqrt{2}} I_{B_0} \cos \omega(t - [(I_3 I_S) \cdot \vec{r}]) + \frac{1}{\sqrt{2}} I_{E_0} \sin \omega(t - [(I_3 I_S) \cdot \vec{r}]) \\
&+ \frac{1}{\sqrt{2}} \cos \omega(t + [(I_3 I_S) \cdot \vec{r}]) + \frac{1}{\sqrt{2}} I_S \sin \omega(t + [(I_3 I_S) \cdot \vec{r}]) \\
&+ \frac{1}{\sqrt{2}} I_{B_0} \cos \omega(t + [(I_3 I_S) \cdot \vec{r}]) + \frac{1}{\sqrt{2}} I_{E_0} \sin \omega(t + [(I_3 I_S) \cdot \vec{r}]) \\
&= \frac{2}{\sqrt{2}} \cos \omega([(I_3 I_S) \cdot \vec{r}]) \left( \cos \omega t + I_S \sin \omega t + I_{B_0} \cos \omega t + I_{E_0} \sin \omega t \right)
\end{aligned}$$

We see that the state $e^{I_{Plane}^+ \varphi^+} + e^{I_{Plane}^- \varphi^-}$ has the form of (1.1) considered in Introduction, thus the results for the Clifford translation obtained there can be applied.

Suppose that the angle in the above Clifford translation continuously changes in time, $\gamma = \omega_r t$. Then Clifford translation with $I_H = I_{E_0}$ and $|H| = 2\gamma$ gives:

$$\begin{aligned}
e^{I_H |H| t} \left( \lambda e^{I_{Plane}^+ \varphi^+} + \mu e^{I_{Plane}^- \varphi^-} \right)\Big|_{\lambda=\mu=1} &= e^{2 I_{E_0} \omega_r t} \left( e^{I_{Plane}^+ \varphi^+} + e^{I_{Plane}^- \varphi^-} \right) \\
&= \cos 2\omega_r t \cos \varphi \\
&- \frac{\beta}{\sqrt{1 - 2\alpha^2}} \sin 2\omega_r t \sin \varphi \\
&+ \left( \frac{\beta}{\sqrt{1 - 2\alpha^2}} \cos 2\omega_r t \sin \varphi + \frac{\alpha}{\sqrt{1 - 2\alpha^2}} \sin 2\omega_r t \sin \varphi \right) B_1 \\
&+ \left( \frac{\alpha}{\sqrt{1 - 2\alpha^2}} \cos 2\omega_r t \sin \varphi + \frac{\beta}{\sqrt{1 - 2\alpha^2}} \sin 2\omega_r t \sin \varphi \right) B_2 \\
&+ \left( \sin \gamma 2\omega_r t \cos \varphi + \frac{\beta}{\sqrt{1 - 2\alpha^2}} \cos 2\omega_r t \sin \varphi \right) B_3
\end{aligned}$$

where $\alpha = \frac{2}{\sqrt{2}} \cos \omega([(I_3 I_S) \cdot \vec{r}]) \cos \omega t$ and $\beta = \frac{2}{\sqrt{2}} \cos \omega([(I_3 I_S) \cdot \vec{r}]) \sin \omega t$

After some trigonometry the result is:

$$e^{2I_{E_0}\omega_r t}\left(e^{I^+_{Plane}\varphi^+} + e^{I^-_{Plane}\varphi^-}\right)$$
$$= \frac{2}{\sqrt{2}}\cos[(I_3 I_S)\cdot\vec{r}]\{\cos(2\omega_r t + \omega t) + \sin(2\omega_r t + \omega t)I_{E_0}$$
$$+ [\cos(2\omega_r t + \omega t) + \sin(2\omega_r t + \omega t)I_{E_0}]I_{B_0}\}$$

We see that action of the Clifford translation $e^{2I_{E_0}\omega_r t}$ on the general solution constructed as equal weight sum of basic solutions of the Maxwell equation gives, up to the space location defined factor, the state which is the sum of rotation in the plane of $I_{E_0}$ by angle $2\omega_r t + \omega t$ plus the same rotation followed by flip around plane of $I_{E_0}$. It can also be read as a single $G_3^+$ element:

$$e^{2I_{E_0}\omega_r t}\left(e^{I^+_{Plane}\varphi^+} + e^{I^-_{Plane}\varphi^-}\right) = 2\cos[(I_3 I_S)\cdot\vec{r}]\, e^{I_R \varphi}$$

where

$$\varphi = \cos^{-1}\left(\frac{1}{\sqrt{2}}\cos(2\omega_r t + \omega t)\right) \tag{2.1}$$

$$I_R = \frac{1}{\sqrt{2}}\sin(2\omega_r t + \omega t)I_S + \frac{1}{\sqrt{2}}\cos(2\omega_r t + \omega t)I_{B_0} + \frac{1}{\sqrt{2}}\sin(2\omega_r t + \omega t)I_{E_0} \tag{2.2}$$

## 3. Measurement of an observable

Though the Clifford translation result received above is linear combination of two geometric algebra states corresponding to some conventional qubits $c_1 \begin{pmatrix}1\\0\end{pmatrix}$ and $c_2 \begin{pmatrix}0\\1\end{pmatrix}$ (see, for example [5]), measurement of a $G_3$ valued observable is not linear combination of measurements by those states because measurement in the currently used formalism does not follow the distributive law. So, we need to use general formula, the geometric algebra variant of the Hopf fibration (see [2], Sec.5.1):

If we have a state $\alpha + \beta_1 B_1 + \beta_2 B_2 + \beta_3 B_3$, $\alpha^2 + \beta_1^2 + \beta_2^2 + \beta_3^2 = 1$, and observable $C_0 + C_1 B_1 + C_2 B_2 + C_3 B_3$ then the result of measurement is:

$$C_0 + C_1 B_1 + C_2 B_2 + C_3 B_3 \xrightarrow{\alpha + \beta_1 B_1 + \beta_2 B_2 + \beta_3 B_3} C_0$$
$$+ \left(C_1[(\alpha^2 + \beta_1^2) - (\beta_2^2 + \beta_3^2)] + 2C_2(\beta_1\beta_2 - \alpha\beta_3) + 2C_3(\alpha\beta_2 + \beta_1\beta_3)\right)B_1$$
$$+ \left(2C_1(\alpha\beta_3 + \beta_1\beta_2) + C_2[(\alpha^2 + \beta_2^2) - (\beta_1^2 + \beta_3^2)] + 2C_3(\beta_2\beta_3 - \alpha\beta_1)\right)B_2$$
$$+ \left(2C_1(\beta_1\beta_3 - \alpha\beta_2) + 2C_2(\alpha\beta_1 + \beta_2\beta_3) + C_3[(\alpha^2 + \beta_3^2) - (\beta_1^2 + \beta_2^2)]\right)B_3$$

In the current case:

$B_1 = I_S$, $B_2 = I_{B_0}$, $B_3 = I_{E_0}$,

$\alpha = \frac{1}{\sqrt{2}}\cos(2\omega_r t + \omega t)$,

$\beta_1 = \frac{1}{\sqrt{2}}\sin(2\omega_r t + \omega t)$, $\beta_2 = \frac{1}{\sqrt{2}}\cos(2\omega_r t + \omega t)$, $\beta_3 = \frac{1}{\sqrt{2}}\sin(2\omega_r t + \omega t)$

and then the result of measurement (without the distance factor $4\cos^2[(I_3 I_S)\cdot\vec{r}]$ ) is:

$$C_0 + C_3 I_S + (C_1 \sin 2(2\omega_r t + \omega t) + C_2 \cos 2(2\omega_r t + \omega t))I_{B_0} +$$

$$(-C_1 \cos 2(2\omega_r t + \omega t) + C_2 \sin 2(2\omega_r t + \omega t))I_{E_0} \qquad (3.1)$$

Geometrically, it means the following:

- The observable in the measurement by $e^{I_R \varphi}$, $\varphi$ and $I_R$ defined in (2.1) and (2.2), is rotated counterclockwise by $\pi/2$ around the axis dual to $I_{B_0}$, thus the observable third bivector component of the value $C_3$ becomes lying in $I_S$ and does not further change;
- Two other components, with initial values $C_1$ and $C_2$ in the planes of $I_S$ and $I_{B_0}$, become orthogonal to $I_S$ and rotate clockwise around axis dual to $I_S$ with angle $\pi/2 - 2(2\omega_r t + \omega t)$[3].

Like in above simple example of Berry potential, the result of measurement, including the factor $4 cos^2[(I_3 I_S) \cdot \vec{r}]$, spreads through the whole space at any instant of time.

## 4. Conclusions

It was shown that for the special case of g-cubit states that particularly appear as solutions of the Maxwell equation in geometric algebra terms the result of measurement of any $G_3$ valued observable is spread through the whole three-dimensional space at any instant of time.

---

[3] Obviously, the result qualitatively does not depend on extra state rotation caused by Clifford translation. The latter just increases the speed of rotation around vector dual to $I_S$.